\documentclass[12pt]{iopart}
\usepackage{iopams}
\usepackage{epsfig}
\bibstyle{plain}
\begin{document}

\title[Emergence of target waves in paced populations of cyclically competing species]{Emergence of target waves in paced populations of cyclically competing species}

\author{Luo-Luo Jiang,$^{1,\ast}$ Tao Zhou,$^{1,2,\dagger}$ Matja{\v z} Perc,$^{3,\ddagger}$ Xin Huang,$^4$ Bing-Hong Wang$^{1,\star}$}

\address{$^1$Department of Modern Physics, University of Science and Technology of China, Hefei 230026, PR China}
\address{$^2$Department of Physics, University of Fribourg, Chemin du Muse 3, CH-1700 Fribourg, Switzerland}
\address{$^3$Department of Physics, Faculty of Natural Sciences and Mathematics, University of Maribor, Koro{\v s}ka cesta 160, SI-2000 Maribor,
Slovenia}
\address{$^4$Department of Physics, University of Science and Technology of China, Hefei 230026, PR China}

\ead{$^\ast$jiangluo@mail.ustc.edu.cn, $^\dagger$zhutou@ustc.edu, $^\ddagger$matjaz.perc@uni-mb.si, $^\star$bhwang@ustc.edu.cn}

\begin{abstract}
We investigate the emergence of target waves in a cyclic predator-prey model incorporating a periodic current of the three competing species in a small area situated at the center of the square lattice. The periodic current acts as a pacemaker, trying to impose its rhythm on the overall spatiotemporal evolution of the three species. We show that the pacemaker is able to nucleate target waves that eventually spread across the whole population, whereby three routes leading to this phenomenon can be distinguished depending on the mobility of the three species and the oscillation period of the localized current. First, target waves can emerge due to the synchronization between the periodic current and oscillations of the density of the three species on the spatial grid. The second route is similar to the first, the difference being that the synchronization sets in only intermittently. Finally, the third route towards target waves is realized when the frequency of the pacemaker is much higher than that characterizing the oscillations of the overall density of the three species. By considering mobility and the frequency of the current as variable parameters, we thus provide insights into the mechanisms of pattern formation resulting from the interplay between local and global dynamics in systems governed by cyclically competing species.
\end{abstract}

\maketitle

\section{Introduction}

Spatially extended excitable systems are investigated frequently due to their omnipresence in several biological, chemical and physical settings \cite{inti1, inti2, inti3, inti4, inti5, inti6}, as reviewed comprehensively in \cite{adr1, adr2}. Especially pattern formation and the related self-organized emergence of target waves have received ample attention in the past \cite{inti7}. Interestingly, although typically lacking an excitable character, systems describing competing interactions in natural and social systems may exhibit similar phenomena as well \cite{inta1, inta2, inta3, inta5, inta6}, including the formation of target waves \cite{intf1, intff2, intf2, intf3, intf4, intf5}. Previous works have revealed that patterns in excitable systems emerge primarily due to the instabilities induced by the interplay between the fast excitatory and the slow recovery variable \cite{inte1, inte2, inte3}. For example, the Bromous acid diffuses much faster than Ferroin in the Belousov-Zhabotinsky reaction, and the cyclic adenosine monophosphate (cAMP) diffuses much faster than membrane receptors in the \textit{Dictyostelium discoideum}. However, many patterns, especially those occurring in mobile populations such as migrating animals or moving bacteria, cannot be explained by the mechanism that is applicable for the emergence of pattern formation in excitable systems. This is mainly because the excitability as a dynamical property is absent in systems where the moving of individuals makes all species equivalent in terms of their diffusion \cite{intd1, intdd1}.

In the social amoeba \textit{Dictyostelium discoideum} periodic currents of cAMP have been introduced in order to study the movement of the cells controlled by the emerging propagating waves of cAMP as during aggregation in the mound \cite{intj1, intj2, intj3}. When \textit{Dictyostelium discoideum} cells were left starving, they begun to emit pulses of cAMP which were relayed to more distant cells by radially divergent waves. Periodic injections of cAMP in slugs were found leading to the chemotactic attraction of anterior-like cells to the tip of the micropipette \cite{intj1, intj2}. Dormann and Weijer reported that propagating chemoattractant waves are thus able to coordinate the periodic movement of the cells, and moreover, that synchronization between the frequency of cAMP injection and cell movement may occur \cite{intj3}. While recently Cyrill \textit{et al.} proposed a model that generates coherent target waves in recurrent single species populations \cite{inti6}, the mechanisms behind the formation of target waves within multi-species populations in the presence of an external periodic current have not yet been investigated.

The cyclic predator-prey model provides a concise description of competition in a multi-species environment. An experimental study by Kerr \textit{et al.} \cite{intd1} revealed that the mechanism of the rock-paper-scissors game can promote biodiversity of the three strains of \textit{Escherichia coli}. Recently, Reichenbach \textit{et al.} proposed a rock-paper-scissors game with mobile players \cite{inta1, intd2}, where the mobility was found to have a critical effect on species diversity. In particular, when mobility was below a critical value all species have been found to coexist by forming spiral waves, whereas above this threshold biodiversity was jeopardized. Motivated by the experimental studies on the three strains of \textit{Escherichia coli} \cite{intd1, intdd1} and the species of \textit{Dictyostelium discoideum} \cite{intj1, intj2, intj3}, we here investigate the pattern formation in a cyclic predator-prey model incorporating a localized periodic current of the three competing species. The periodic current essentially acts as a pacemaker on the population, trying to impose its rhythm on the spatiotemporal evolution of the species, but it can also be regarded as a localized inhomogeneity affecting the overall system dynamics \cite{con1}. We report the emergence of coherent target waves, in general emerging due to the nucleation induced by the localized current. More precisely, three routes to target wave formation can be distinguished based on the interplay between local and global dynamics. Local thereby refers to the properties associated with the oscillations within the paced region of the spatial grid whereas global refers to the overall dynamics of the three species. In particular, target waves can emerge either due to the constant or due to the intermittent synchronization between the local and global dynamics, or when the frequency of the pacemaker, reflecting directly the so-called local dynamics, is much higher than the frequency characterizing the oscillations of the overall density of the three species, \textit{i.e.} the global dynamics. Interestingly, although the presently employed model is based on microscopic interactions among individuals only, the target waves emerge on a macroscopic level. Furthermore, our work indicates possibilities of pattern control and selection in systems governed by cyclical interactions.

\section{Three-species cyclic predator-prey model}

Based on previous works of Reichenbach \textit{et al.} \cite{inta1, intc1, intd2}, we employ a three-species cyclic predator-prey model with the following specifications. Nodes of a $L \times L$ square lattice present mobile individuals belonging to one of the three species, which we denote by $A$, $B$ and $C$. Each node can either host one individual of a given species or it can be vacant. Vacant sites, which we denote by $\otimes$, are also the so-called resource sites. Within the model three processes are possible; namely predation, reproduction and exchange, whereby these occur only between neighboring nodes. $\emph{Predation.}$-- Species $A$ eliminates species $B$ at a rate $1$, whereby the node previously hosting species $B$ becomes vacant. In the same manner species $B$ can eliminate species $C$, and species $C$ can eliminate species $A$, thus forming a closed loop of dominance between them. $\emph{Reproduction.}$-- Individuals can place an offspring to a neighboring vacant node $\otimes$ at a rate $1$. $\emph{Exchange.}$-- Two individuals, including vacant sites, can exchange their position at a rate $\alpha$, thus introducing mobility of participants. According to this description, predation (upper row), reproduction (middle row) and exchange (lower row) can be described by the reactions:
\begin{eqnarray}
AB \stackrel{1}{\longrightarrow} A\otimes&\,, \quad
BC \stackrel{1}{\longrightarrow} B\otimes&\,, \quad
CA \stackrel{1}{\longrightarrow} C\otimes\, \cr
A\otimes \stackrel{1}{\longrightarrow} AA&\,, \quad
B\otimes \stackrel{1}{\longrightarrow} BB&\,, \quad
C\otimes \stackrel{1}{\longrightarrow}CC\, \cr
XY \stackrel{\alpha}{\longrightarrow} YX&\,\nonumber
\end{eqnarray}
where $X,~Y \in \{A,B,C, \otimes\}$. According to the random walk theory \cite{stoa3}, the mobility of individuals $M$ can be defined in terms of the exchange process as $M=2 \alpha / L^2$, meaning it is proportional to the typical area explored by a mobile individual per unit time.

Unlike the deterministic approach, which regards the time evolution as a continuous process, here we make use of a stochastic simulation algorithm whereby the temporal evolution can be considered as a random walk process. The most commonly applied stochastic simulation algorithm was developed by Gillespie \cite{stoa1, stoa2}, where reactions occur in a random manner. In particular, predation and reproduction occur with probability $1 /(\alpha +2)$, whereas exchange (moving) occurs with probability $\alpha /(\alpha + 2)$. If one ignores the spatial structure and assumes the system to be well mixed, the model can be described by partial differential equations. The results presented below, however, are obtained via Monte Carlo simulations of the $L \times L$ square lattice, whereby we use no-flux boundary conditions and the algorithm of Gillespie \cite{stoa1, stoa2} for determining the probabilities for the three possible processes. An elementary Monte Carlo step consists of randomly choosing an individual who interacts with one of its four nearest neighbors, which is also selected randomly, and then executing the process as determined by the Gillespie's algorithm. One full Monte Carlo step consists of $N=L^2$ elementary steps, during which, in accordance with the random sequential update, each player is selected once on average. The resolution of time is thus measured in full Monte Carlo steps.

\begin{figure}
\begin{center} \includegraphics[width = 7cm]{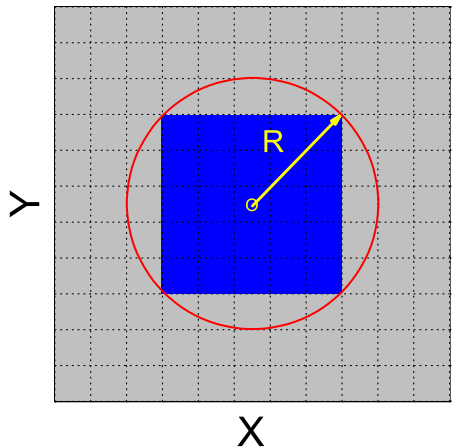}
\caption{\label{injection} (Color online) Schematic presentation of the periodic current introduced to an $11 \times 11$ lattice. Here the radius is equal to $R=3.5$. The blue area is subjected to a periodic injection of the three species, in the recurrent order $A \rightarrow B \rightarrow C \rightarrow A$, each every multiple of $T_{0}$ (see also main text for details). The gray area is initially populated either by $A$, $B$, $C$ and $\otimes$ with equal probability [initial condition 1 (IC1)] or solely with vacant sites [initial condition 2 (IC2)]. Throughout this paper we will use $R=10.5$ and $L$ such that $0.01 \leq R/L \leq 0.02$ unless stated otherwise. We note that the results reported below are robust to variations of this ratio as long as the pacing area (blue) is large enough to nucleate target waves by appropriate $M$ and $T_{0}$.}
\end{center}
\end{figure}

Furthermore, motivated by the application of periodic currents of cAMP in the experiments with \textit{Dictyostelium discoideum} \cite{intj1, intj2, intj3}, we introduce a localized periodic current of the three competing species, as described and schematically presented in Fig.~\ref{injection}. In particular, the periodic current is applied over a small (compared to the overall system size) area located at the center of the spatial grid. The periodic current thus acts as a pacemaker on the population, trying to impose its rhythm on the spatiotemporal evolution of the three species. The current is defined as follows: at time $t=0$ the nodes inside $R$ (see Fig.~\ref{injection}) are populated by species $A$, at time $t=T_{0}$ these nodes are populated by species $B$, at time $t=2T_{0}$ these nodes are populated by species $C$, and at time $t=3T_{0}$ these nodes are again populated by species $A$, and continuing further in this manner. Importantly, during $n T_{0}<t<m T_{0}$, where $n \in \mathbb{Z}^+$ and $m=n+1$, the evolution of species inside $R$ (blue area in Fig.~\ref{injection}) is governed by the same Monte Carlo updating as outside (gray area in Fig.~\ref{injection}). Moreover, from the definition of the current it follows that its period equals $T_{in}=3T_{0}$. In the following, we will consider the mobility $M$ and the time $T_{0}$ between successive replacements of a species inside $R$ as the two crucial parameters effecting the emergence of target waves in the examined model.

Before we start presenting the results, we note that in the absence of the periodic current the model returns identical results in terms of the coexistence of species in dependence on $M$ as reported earlier by Reichenbach \textit{et al.} \cite{inta1}. In particular, while low $M$ ensure coexistence of all three species, there exists a critical $M \approx 10^{-3}$ beyond which diversity is no longer sustained.

\section{Emergence of coherent target waves}

\begin{figure}
\begin{center} \includegraphics[width = 12cm]{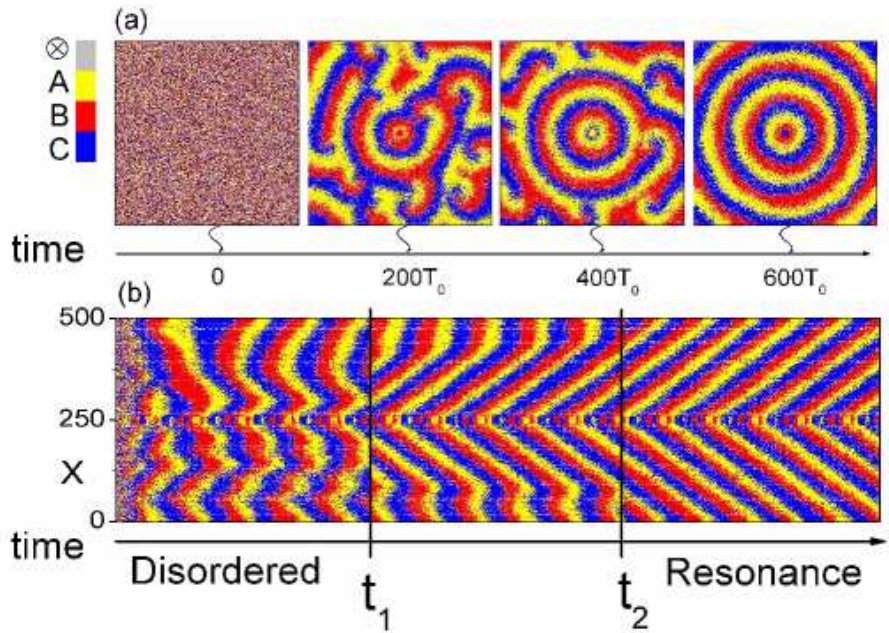}
\caption{\label{coherent} (Color online) (a) Snapshots depicting a typical emergence of target waves in the examined system over time. (b) Space-time plot corresponding to the snapshots in panel (a), obtained at $Y=250$ (cross-section of the $Y$ axis). To ensure clarity some frames within $t_{1} \leq t \leq t_{2}$ were discarded; $t_{1}$ and $t_{2}$ corresponding to $33T_{0}$ and $466T_{0}$, respectively. In both panels yellow, red and blue squares denote species $A$, $B$ and $C$, respectively, whereas the grey squares depict vacant sites. Employed parameter values are: $M=10^{-4}$, $T_{0}=150$ and $L=500$. Initial conditions: IC1 (see the caption of Fig.~\ref{injection}). The full spatiotemporal evolution of the system is recorded in the file \textit{video2.avi}, which is part of the supplementary material pertaining to this paper.}
\end{center}
\end{figure}

We start by presenting snapshots of the spatial grid obtained at different times $t$ in Fig.~\ref{coherent}(a). It is evident that the initial random configuration ($t=0$) is gradually replaced by supremely ordered target waves ($t=600T_{0}$), which emerge after a transient period that is characterized by disordered turbulence at intermediate times ($t=200T_{0}$ and $t=400T_{0}$). The emergence of target waves depicted in the rightmost panel of Fig.~\ref{coherent}(a) is a direct consequence of the localized periodic current, \textit{i.e.} the pacemaker, which periodically introduces an inhomogeneous nucleus at the center of the spatial grid. These inhomogeneities, in form of regions dominated by a single species, spread and eventually self-organizes into a beautiful manifestation of pattern formation in a population governed by cyclical interactions. A more precise temporal evolution of target wave formation in the system can be visualized by means of the corresponding space-time plot, as is shown in Fig.~\ref{coherent}(b). From the latter it is indeed evident that the target waves begin to grow around the impact site of the periodic current (centered at $X=Y=250$; note that $L=500$), which periodically introduces inhomogeneous regions governed by a single species. Following a period of disordered patters and very turbulent waves ($t<t_{1}$), as well as an intermediate period during which target waves gradually begin to dominate ($t_{1} \leq t \leq t_{2}$), the complete dominance of ordered target waves finally prevails ($t>t_{2}$) in a resonance-like manner due to the interplay between the localized periodic current and the overall spatiotemporal dynamics of the three species. It is worth noting that the two times $t_{1}$ and $t_{2}$ in Fig.~\ref{coherent}(b), denoting the start of target waves formation and their complete dominance, respectively, depend somewhat on the random realization of initial conditions and the location of the periodic current. However, such deviations are negligible and do not affect our results vitally.

\begin{figure}
\begin{center} \includegraphics[width = 13cm]{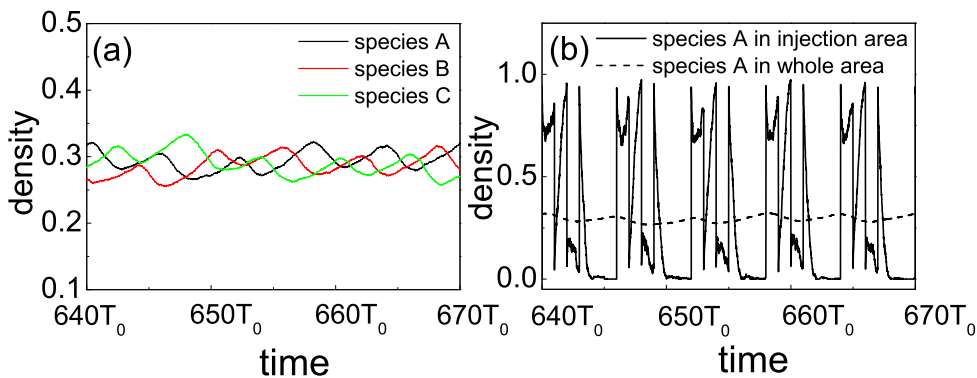}
\caption{\label{oscillation} (Color online) (a) Oscillations of the overall density of the three species on the spatial grid after coherent target waves start dominating, \textit{i.e.} $t>t_{2}$. (b) Comparison of the local (solid line) and global (dashed line) density of species $A$ evolving in time. Note the synchronization between the two depicted quantities (see also main text for details). Employed parameter values and initial conditions are the same as in Fig.~\ref{coherent}.}
\end{center}
\end{figure}

Further elaborating on the emergence of target waves depicted in Fig.~\ref{coherent}, we present in Fig.~\ref{oscillation}(a) oscillations of the overall density of the three species after coherent target waves start dominating the spatial grid. In accordance with the spatiotemporal dynamics of target waves, the temporal outlays of the density of all three species are predominantly periodic, which manifests as sharp peaks in the corresponding Fourier transformation of the series, occurring at multiples of the main oscillation frequency. It is more instructive, however, to compare the oscillations of the density of a given species measured only within the paced (injection) area and its overall density on the spatial grid. Figure~\ref{oscillation}(b) features these two temporal plots for species $A$, and it can be observed that the two are in synchrony. This confirms that the target waves emerge as a consequence of the interplay between local (encompassing only the paced area) and global (encompassing the whole spatial grid) system dynamics, which during the dominance of ordered target waves [see rightmost panel of Fig.~\ref{coherent}(a)] can be synchronized.

\begin{figure}
\begin{center} \includegraphics[width = 15cm]{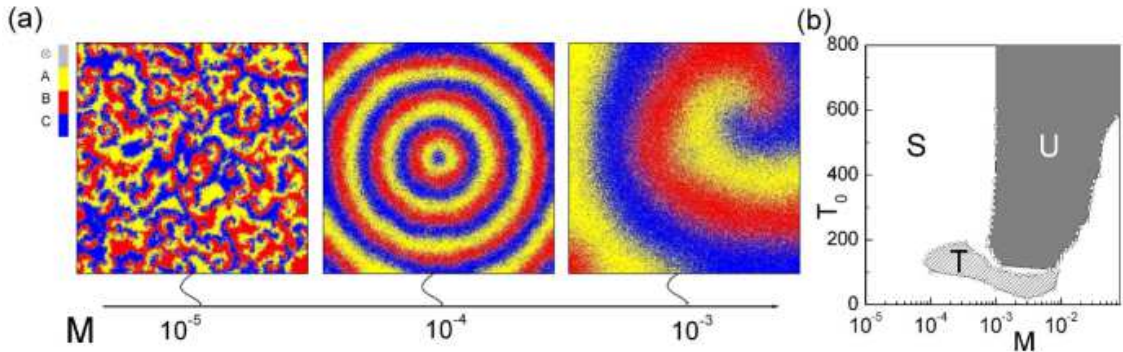}
\caption{\label{phase} (Color online) (a) Characteristic snapshots of the spatial grid for $M=10^{-5}$ (left), $M=10^{-4}$ (middle) and $M=10^{-3}$ (right), obtained after long transients have been discarded. In all three panels yellow, red and blue squares denote species $A$, $B$ and $C$, respectively, whereas the grey squares depict vacant sites. Employed parameter values are: $T_{0}=150$ and $L=500$. Initial conditions: IC1 (see the caption of Fig.~\ref{injection}). For the full spatiotemporal evolution of the system see also the supplementary files \textit{video1.avi} ($M=10^{-5}$), \textit{video2.avi} ($M=10^{-4}$) and \textit{video3.avi} ($M=10^{-3}$). (b) Phase diagram revealing the evolutionary outcomes of the system depending on the combination of $M$ and $T_{0}$. White region (S) indicates the formation of turbulent or ordered spiral waves [left and right panel in (a)], hatched region (T) indicates the formation of coherent target waves [middle panel in (a)], and the gray region (U) indicates a uniform final state that is dominated by a single randomly selected species.}
\end{center}
\end{figure}

Nevertheless, above results pertain to a single combination of $M$ and $T_{0}$, and thus it is of interest to investigate the impact of other values of these two parameters on pattern formation as well. Figure~\ref{phase}(a) shows typical snapshots of the spatial grid obtained for different mobility $M$ of individuals after a long simulation time. Evidently, low mobility fails to evoke target patterns. In the leftmost panel of Fig.~\ref{phase}(a), where $M=10^{-5}$, turbulent spirals dominate over the entire lattice. For $M=10^{-4}$, on the other hand, the locally introduced inhomogeneities due to the periodic current are appropriately enhanced to eventually result in the emergence of target waves, as depicted in the middle panel of Fig.~\ref{phase}(a). Increasing the mobility further to $M=10^{-3}$ evokes a transition from target to spiral waves, as depicted in the rightmost panel of Fig.~\ref{phase}(a), which is due to the strong mixing of individuals prohibiting the stability of the relatively stationary target wave pattern and instead favoring the more dynamically evolving spiral waves. Full spatiotemporal evolutions of the system are recorded in the supplementary files \textit{video1.avi}, \textit{video2.avi} and \textit{video3.avi} pertaining to this paper, for $M=10^{-5}$, $M=10^{-4}$ and $M=10^{-3}$, respectively.

The impact of different values of $M$ and $T_{0}$ can be captured succinctly by the phase diagram presented in Fig.~\ref{phase}(b), which results from a systematic study of pattern formation over the whole 2D parameter plane. Depending on the combination of $M$ and $T_{0}$, three different evolutionary outcomes can be distinguished. Namely, the formation of coherent target waves (region T), the formation of turbulent or ordered spiral waves (region S), or the formation of a uniform phase (region U) following extinction of two species. From the phase diagram presented in Fig.~\ref{phase}(b) it follows that the region of coherent target waves is relatively small, and that thus the phenomenon results from a rather subtle interplay between the localized periodic current and the overall dynamics of the three species. It is worth noting that these results are largely independent of the system size as long as the ratio $R/L$ remains the same, which ensures that the local dynamics induced by the periodic current is sufficiently represented in the system. Obviously, increasing the value of $R$ by a given system size $L$ facilitates the formation of target waves, whereas smaller $R$ are less likely to have a noticeable impact on the overall evolution of the three species. In general, however, quantitative differences due to such variations are small, affecting the borders of the hatched region (T) [Fig.~\ref{phase}(b)] in the $M - T_{0}$ plane slightly. We also note that the border separating turbulent and ordered spiral waves in the white region (S) is located at $M = 6 \cdot 10^{-5}$. We have observed that smaller values of $M$ lead to turbulent spiral waves [see the leftmost panel of Fig.~\ref{phase}(a)], while larger values of $M$ within the S region tend to yield ordered spiral waves [see the rightmost panel of Fig.~\ref{phase}(a)].

\section{Three routes towards target waves}

In order to develop a better understanding of the emergence of coherent target waves evoked by the localized periodic current for different values of $M$ and $T_{0}$, we change the initial setup in that we use IC2 type initial conditions, \textit{i.e.} apart from the paced area, the lattice is initially populated solely with vacant sites (see also the caption of Fig.~\ref{injection}). This is inspired by the growth experiments of \textit{Escherichia coli} \cite{intff2,intf2}, and enables us a clearer tracking of the localized nucleus formation and subsequent spreading of the periodically injected species across the spatial grid. In the following, we will show that indeed three routes to the emergence of coherent target waves can be distinguished.

\begin{figure}
\begin{center} \includegraphics[width = 16cm]{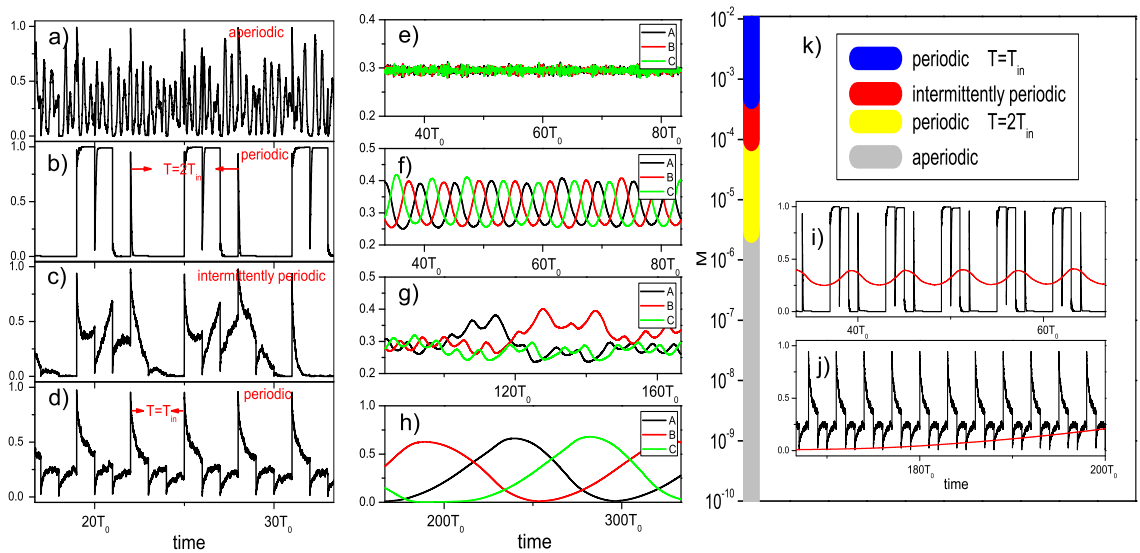}
\caption{\label{proportion} (color online) Panels (a)-(d) feature temporal evolutions of the density of species $A$ within the area that is subjected to the periodic current for different values of $M$ ($10^{-6}$, $10^{-5}$, $10^{-4}$ and $10^{-3}$ from top to bottom). Panels (e)-(h) feature temporal evolutions of the overall density of the three species on the spatial grid for different values of $M$ ($10^{-6}$, $10^{-5}$, $10^{-4}$ and $10^{-3}$ from top to bottom). Panels (i) and (j) depict the oscillations of species $A$ within the paced area (black line) and the overall density of species $A$ (red line) jointly for $M=10^{-5}$ and $M=10^{-3}$, respectively. Panel (k) shows system states in terms of the oscillatory properties from $M=10^{-10}$ to $M=10^{-2}$. For all panels the employed parameter values are: $T_{0}=600$ and $L=1024$. Initial conditions: IC2 (see the caption of Fig.~\ref{injection}).}
\end{center}
\end{figure}

\begin{figure}
\begin{center} \includegraphics[width = 13cm]{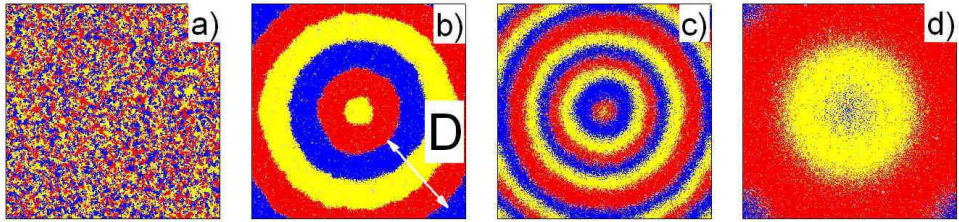}
\caption{\label{target} (color online) Characteristic snapshots of the spatial grid for $M=10^{-6}$ (a), $M=10^{-5}$ (b), $M=10^{-4}$ (c) and $M=10^{-3}$ (d), corresponding to the oscillatory traces depicted in Figs.~\ref{proportion}(a,e), (b,f), (c,g) and (d,h), respectively. Other employed parameter values and initial conditions are thus the same as in Fig.~\ref{proportion}.}
\end{center}
\end{figure}

First, we present the temporal evolution of the density of species $A$ within the area that is subjected to the periodic current for different values of $M$ in Figs.~\ref{proportion}(a-d). It can be observed that the outlay for $M=10^{-6}$ [panel (a)] is aperiodic, indeed resembling chaotic oscillations (this can be confirmed by calculating the maximal Lyapunov exponent \cite{hk1}, which for this case is positive), becomes periodic for $M=10^{-5}$ [panel (b); oscillation period is $T=2T_{in}=3600$] and $M=10^{-3}$ [panel (d); oscillation period is $T=T_{in}=1800$], and exhibits intermittent episodes of periodicity for $M=10^{-4}$ [panel (c)]. Figures~\ref{proportion}(e-h) feature the temporal outlay of the overall density of the three species on the spatial grid for the corresponding values of $M$ used in panels (a-d). Finally, Fig.~\ref{proportion}(i) and Fig.~\ref{proportion}(j) depict the oscillations of species $A$ within the paced area (black line) and the overall density of species $A$ (red line) jointly for $M=10^{-5}$ and $M=10^{-3}$, respectively. Importantly, while for $M=10^{-5}$ the two depicted time course are synchronized, as is the case also in Fig.~\ref{oscillation}(b), for $M=10^{-3}$ the oscillation period of the overall density of species $A$ on the spatial grid is much larger than that of the density within the paced area. Thus, the two cases constitute two different routes towards target waves [see also Figs.~\ref{target}(b) and (d) for the corresponding snapshots] in the examined system. The third route emerges if $M=10^{-4}$, where the synchronization between local and overall oscillations occurs only intermittently. As a consequence, the resulting target waves do not have a well-defined wavelength, as can be observed in Fig.~\ref{target}(c). Figure~\ref{proportion}(k) depicts succinctly the intervals of $M$ (for the considered $T_{0}=600$) resulting in a given route to coherent target formation. Notably, the gray region of $M$ (aperiodic regime) results only in disordered turbulent spiral-like waves [see also Fig.~\ref{target}(a) for the corresponding snapshot].

From the results presented in Fig.~\ref{proportion}, we thus conclude that there exist three different routes to coherent target waves in the examined cyclic predator-prey model incorporating a periodic current. First, target waves can emerge due to the synchronization between the periodic current and oscillations of the density of the three species on the spatial grid, as presented in Fig.~\ref{proportion}(i). The resulting spatial portrait is shown in Fig.~\ref{target}(b). Due to the regularity of the oscillations the target waves also have an easily inferable wavelength $\lambda=D/L$, as noted on the corresponding snapshot. The second route is similar to the first, the difference being that the synchronization sets in only intermittently. The resulting spatial portrait is shown in Fig.~\ref{target}(c), whereby the seeming lack of a well-defined wavelength is a direct consequence of the intermittent temporal outlay of the variations of the species' densities, as shown in Figs.~\ref{proportion}(c) and (g). Note that in Fig.~\ref{target}(c) the circular ring of the blue species, for example, is sometimes thinner and sometimes thicker, \textit{i.e.} not always the same. This is because the oscillation period of the underlying temporal trace is not exact but varies intermittently. Finally, the third route towards target waves is realized when the frequency of the pacemaker is much higher than that characterizing the oscillations of the overall density of the three species, as presented in Fig.~\ref{proportion}(j). The resulting spatial portrait is shown in Fig.~\ref{target}(d), whereby the comparably very large $\lambda$, especially if compared to the $\lambda$ of the snapshot presented in panel (b), is directly linked to the large oscillation period of the overall density of the three species [see Fig.~\ref{proportion}(h) and compare to panel (f)]. It is also worth noting that for sufficiently large $M$, exceeding $5.0\times 10^{-4}$, the mobility quickly decimates any given species that is injected within the paced area, and thus the artificially established dominance there is very short-lived [note the sharp peaks in Fig.~\ref{proportion}(d)]. In fact, individuals of an injected species always aggregate near the boundary of the paced area, and only then start to spread outwards across the spatial grid. But since this aggregation and propagation takes a long time, the period of overall oscillations is much longer than that of oscillations within the paced area. On the other hand, for extremely small values of $M$ ($<3.0 \times 10^{-6}$), individuals spreading out of the area that is subjected to the periodic current distribute almost homogeneously across the spatial grid due to the relatively high rate of predation and reproduction if compared to the exchange rate (mobility). Thus, the injected individuals become fragmented, in turn failing to evoke ordered target waves, as depicted in Fig.~\ref{target}(a). Accordingly also, the oscillations of the species' densities, both within the paced area and overall, are aperiodic [see Figs.~\ref{proportion}(a) and (e)].

\begin{figure}
\begin{center} \includegraphics[width = 14cm]{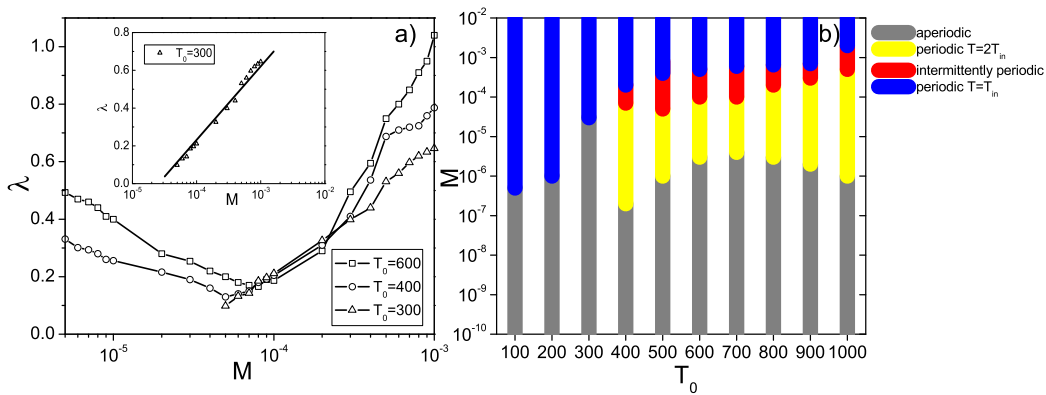}
\caption{\label{wavelength} (color online) (a) Wavelengths $\lambda$ of target waves in dependence on $M$ for different values of $T_{0}$. The inset shows
$\lambda \sim \log M^{1/2}$. (b) Different system states in terms of the oscillatory properties in the $M-T_{0}$ parameter space. Note that the latter uniquely determines the route to coherent target waves in the examined system (see also main text for details).}
\end{center}
\end{figure}

Focusing further on the wavelength $\lambda$ of the target waves depicted in Fig.~\ref{target} (where applicable), one can thus observe that the mobility very much affects the typical spatial distance between neighboring wave fronts with the same species, and while the first [panel (b)] and the third route [panel (d)] yield target waves with an easily inferable $\lambda$, this is not necessarily the case for the second route [panel (c)]. In fact, results presented in Fig.~\ref{wavelength}(a) show that the wavelength of target waves displays non-monotonous behavior in dependence on the mobility of individuals for $T_{0}=600$ and $T_{0}=400$, while increases with $M$ according to $\lambda \sim \log M^{1/2}$ for $T_{0}=300$. Notably, this is different from what has been reported for spiral waves \cite{inta1}, where $\lambda \sim M^{1/2}$ has been found applicable. Furthermore, results in Fig.~\ref{wavelength}(a) suggest that indeed both, $M$ and $T_{0}$ affect the route towards the emergence of target waves. Figure~\ref{wavelength}(b) presents results of an extensive analysis performed over the $M-T_{0}$ parameter space, in particular indicating the properties of the oscillations of species' densities, similarly as in Figure~\ref{proportion}(k). Gray color denotes that, for the particular combination of $M$ and $T_{0}$ values, the emergence of coherent target waves is impossible due to an aperiodic outlay of the oscillations, both within the paced area as well as overall [see Figs.~\ref{proportion}(a) and (e) for an example]. Yellow color indicates the emergence of coherent target waves via the first route, which is due to the synchronization between the periodic current and oscillations of the overall density of the three species on the spatial grid [see Fig.~\ref{proportion}(i) for an example]. Red color indicates the emergence of coherent target waves via the second route (the synchronization occurs intermittently) [see Figs.~\ref{proportion}(c) and (g) for an example], and the blue color indicates the applicability of the third route [see Fig.~\ref{proportion}(f) for an example]. From the results presented in Fig.~\ref{wavelength}(b) it is also clear that the non-monotonous dependence of $\lambda$ on mobility for $T_{0}=400$ and $T_{0}=600$ depicted in Fig.~\ref{wavelength}(a) is due to the emergence of target waves via the second route for intermediate $M$ (red stripes). Finally, we note that when the route towards the emergence of target waves is switched, narrow parameter regions of spiral wave formation may appear [see also Fig.~\ref{phase}(b)], which, interestingly, were also observed in the \textit{Dictyostelium discoideum} studies \cite{inti6, con1a}.

\section{Summary}

In sum, we have studied the emergence of target waves in a cyclic predator-prey model incorporating a periodic current of the three competing species in a small area situated at the center of the square lattice. We have shown that a pacemaker-like periodic current may evoke coherent target waves, provided the mobility of individuals on the spatial grid and the frequency of the forcing are adequately adjusted. Furthermore, we have identified three possible routes towards the emergence of coherent target waves, depending on the properties of the oscillations of species' densities within the paced area and across the whole spatial grid. In particular, target waves can emerge due to the complete or intermittent synchronization between the periodic current and oscillations of the overall densities of the three species, or they can emerge so that the frequency of the pacemaker is much higher than the frequency characterizing the induced global oscillations of the three densities. Irrespective of the scenario, however, the global oscillations and the associated target waves are a direct consequence of the introduced pacemaker. Which route is chosen depends both on the mobility of the individuals as well as on the selected oscillation period of the localized current. We thus provide insights into the mechanisms of pattern formation resulting due to the interplay between local and global dynamics in a system that is governed by cyclically competing species. Since the identified mechanisms affect the spatiotemporal dynamics in a predictable way, and moreover, can be tuned effectively via accessible system parameters like the frequency of the pacemaker, our work indicates possibilities of pattern control and selection in systems governed by cyclical interactions. Cyclical competitions evidently play an important role by the maintenance of ecological biodiversity and emergence of cooperation in structured populations \cite{conb1, conb2, conb3, s1, s3, cfx1}. Related to this, complex spatiotemporal formations have often been observed, as for example spiral wave structures in the realm of the rock-paper-scissors game \cite{conb8}. Coherent target waves, however, are rarely reported in this context. Although target waves were observed in excitable systems, the main difference of the present model is that the diffusion induced by individual mobility is the same for all species. Therefore, the emergence of target waves is due to a different mechanism. Notably, the periodic injection method has recently also been used in a complex Ginzburg-Landau system for the purpose of spatiotemporal chaos control \cite{con1, con2}. According to our study, one can control the route towards target wave formation by adjusting the frequency of the periodic current. In this way also the wavelength of target waves can be controlled, which may find useful applications in a variety of realistic systems, ranging from targeted drug delivery to sustenance of biodiversity. We hope that our study will promote the understanding of social dynamics and strengthen the established importance of physics \cite{hij} when striving towards this goal.

\ack{This work is supported by the National Natural Science Foundation of China (Grant Nos. 10635040 and 70871082), and by the Specialized Research Fund for the Doctoral Program of Higher Education of China (SRFD No. 20070420734). Matja{\v z} Perc additionally acknowledges support from the Slovenian Research Agency (Grant No. Z1-2032-2547).}

\end{document}